\documentclass[conference]{IEEEtran}
\IEEEoverridecommandlockouts

\usepackage[left=1.62cm,right=1.62cm,top=1.9cm]{geometry}
\usepackage{cite}
\usepackage{amsmath,amssymb,amsfonts}
\usepackage{algorithmic}
\usepackage{graphicx}
\usepackage{textcomp}
\usepackage{amsmath}
\DeclareMathOperator{\diag}{diag}
\usepackage{amsmath}
\usepackage{mathtools}
\usepackage{verbatim}
\usepackage{xcolor}

\usepackage[T1]{fontenc}
\usepackage[utf8]{inputenc}
\usepackage{authblk}

\def\BibTeX{{\rm B\kern-.05em{\sc i\kern-.025em b}\kern-.08em
    T\kern-.1667em\lower.7ex\hbox{E}\kern-.125emX}}
\begin{document}

\title{Outage Analysis for Active Reconfigurable Intelligent Surface-Enhanced Wireless Powered Communication Networks\\

{
} 
\thanks{This work was supported by Basic Science Research Program through NRF
funded by the MOE (NRF-2022R1I1A1A01071807, 2021R1I1A3041887), and
Institute of Information \& communications Technology Planning \& Evaluation
(IITP) grant funded by the Korea government (MSIT) (2022-0-00704, and Development
of 3D-NET Core Technology for High-Mobility Vehicular Service).}
}

\author[*]{Waqas Khalid}
\author[$\dagger$]{Heejung Yu}
\author[$\ddagger$]{Alexandros-Apostolos A. Boulogeorgos}
\affil[*]{Institute of Industrial Technology, Korea University, Sejong, Korea; waqas283@\{korea.ac.kr, gmail.com\}}
\affil[$\dagger$]{Dept. of Elec. \& Inform. Eng., Korea University, Sejong, Korea; heejungyu@korea.ac.kr}
\affil[$\ddagger$]{Dept. of ECE, University of Western Macedonia, Kozani, Greece; al.boulogeorgos@ieee.org}
\renewcommand\Authands{ and }




\maketitle

\begin{abstract}
Wireless powered communication (WPC) involves the integration of energy harvesting and data transmission. This allows devices to communicate without constant battery replacements or wired power sources. Reconfigurable intelligent surfaces (RISs) can dynamically manipulate radio signals. In this paper, we explore the use of active elements to mitigate double-fading challenges inherent in RIS-aided links. We enhance the reliability performance for an energy-constrained user by combining active RIS and WPC. The theoretical closed-form analysis, which includes transmission rate, harvested energy, and outage probability, provides valuable insights that inform parameter selection.
\end{abstract}

\begin{IEEEkeywords}

Wireless powered communication (WPC), Active RIS, Outage probability.
\end{IEEEkeywords}

\section{Introduction}
In the realm of sixth-generation (6G) wireless networks, the deployment of reconfigurable intelligent surfaces (RIS) stands as a pivotal innovation, shaping the foundation of intelligent wireless environments, alleviating radio frequency (RF) complications, and bolstering connectivity under challenging propagation conditions. By introducing controlled phase shifts to electromagnetic signals, RIS systems substantially enhance network coverage and reliability without complex signal processing or additional infrastructure\cite{b1,b2}. Wireless powered communication (WPC) represents another emerging field anticipated to play a pivotal role in 6G networks, seamlessly merging energy harvesting with data transmission. This integration empowers networked devices to communicate independently of conventional power sources, eliminating dependence on batteries or a constant power supply \cite{b3}. 

\subsection{Motivation}
In practical scenarios, the deployment of RIS-aided wireless communication systems may confront challenges, including the significant issue of double-fading \cite{b4,b4a}. The strategic incorporation of active elements in RIS has been proposed, demonstrating considerable potential in effectively overcoming the double-fading challenge \cite{b5}. The synergistic amalgamation of active RIS with WPC emerges as a highly promising strategy, substantially enhancing both wireless reliability and energy efficiency, particularly for users facing energy limitations. This innovative integration is poised to significantly elevate network performance, energy efficiency, and connectivity, fostering a new era of advanced, sustainable, and intelligent wireless communications. The deployment of active RIS-WPC in 6G networks holds immense potential, particularly in revolutionizing sectors such as the Internet of Things (IoT), wearable technology, and autonomous vehicles \cite{b6}.

\subsection{Contribution}
In our paper, we focus on RIS-WPC networks, where we employ RIS with active elements to enhance information transmission. Specifically, an energy-constrained wireless user harvests RF energy and employs it for data transmission. This approach capitalizes on RIS beamforming gain to optimize information reception and provides superior adaptability compared to conventional WPC systems without RIS. Our work presents significant contributions including a detailed statistical channel characterization that accounts for potential noise in active RIS modules, a comprehensive closed-form outage probability analysis specific to Rayleigh fading channels, and numerical results. These results not only validate our theoretical findings but also offer essential guidance for parameter selection.

\begin{figure}[t]
\centering
\includegraphics[width=3.4in,height=3in]{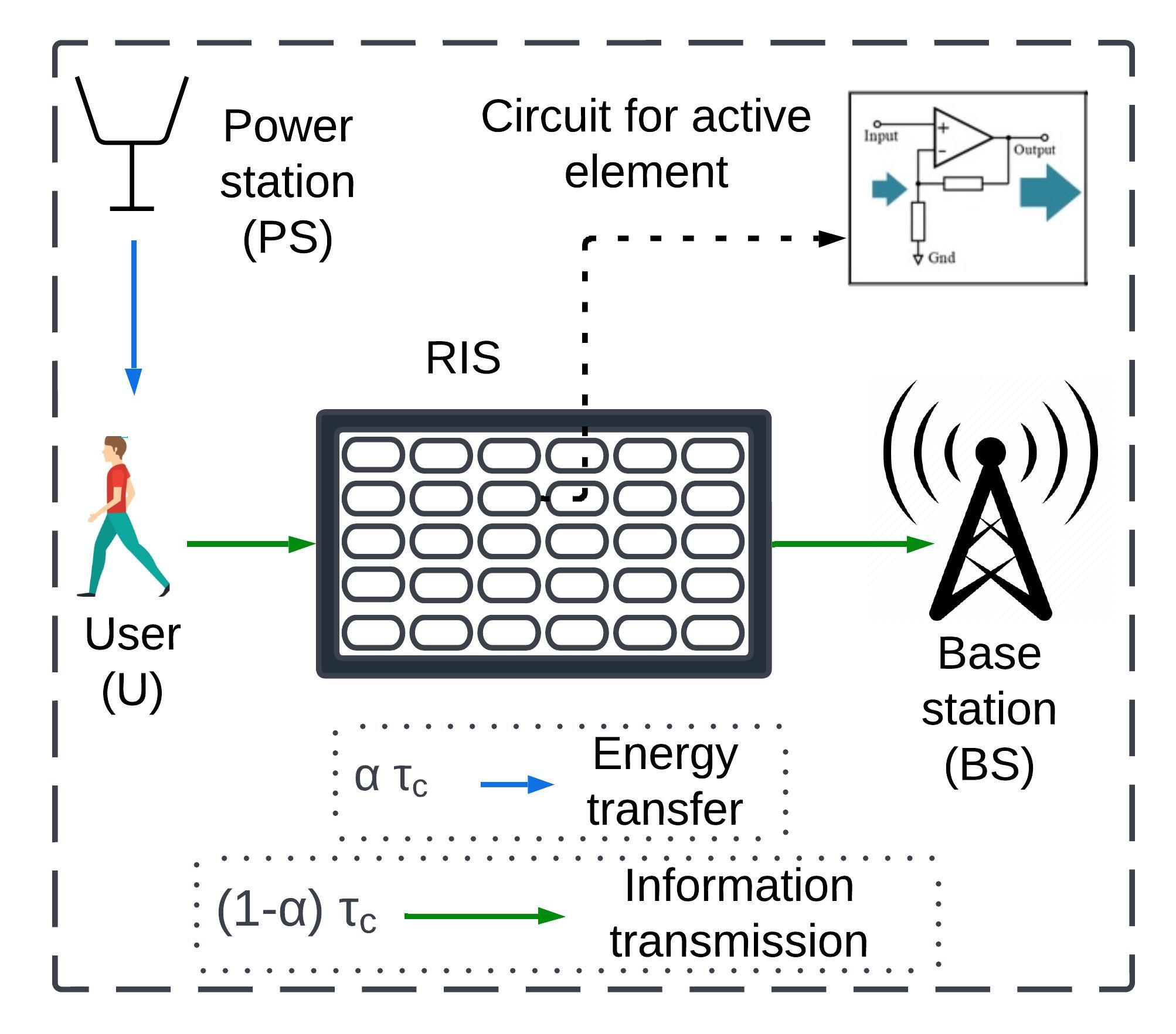}
\caption {Active RIS-Enhanced WPC Network.}
\label{diagram1}
\end{figure}

\section{System Model}
We examine a  WPC network, depicted in Fig. \ref{diagram1}, consisting of a base station (BS), a power station (PS), and a user (U). We introduce an RIS with $M$ elements strategically placed between the BS and U because a direct link is obstructed. Each element within the RIS module is equipped with a power amplifier to amplify the reflected signal. In this configuration, it becomes possible to mitigate the substantial path loss associated with the RIS-aided link, effectively overcoming the challenge posed by double-fading\footnote{Double-fading can be mitigated by increasing passive RIS elements. However, this may result in an increased physical size of the RIS module.}. We adopt an independent and identically distributed (i.i.d.) Rayleigh fading model for all links \cite{b7}. The channel vectors are defined as follows:

\begin{itemize}
  \item $h_1$ for the PS-U link,
  \item  $\textbf{h}_{2}$ for the U-RIS link,
  \item $\textbf{h}_{3}$ for the RIS-BS link. 
\end{itemize}

We consider $h_1$ and each entry in $\textbf{h}_{2}$ and $\textbf{h}_{3}$ to be an independent complex Gaussian distribution with a zero mean and a unit variance. Within each channel coherence interval ($\tau_c$), communication unfolds in two distinct phases: energy transfer and information transmission. During the $\alpha \tau_c$ duration, the PS facilitates the transfer of RF energy to $U$. 

The energy harvested by U can be expressed as follows:

 \begin{align}
 E=\eta  \alpha \tau_c P_{ps}  |h_1|^2 \zeta_1
  \end{align}
where 
\begin{itemize}
   \item $0<\alpha<1$ is time switching factor between energy transfer and information transmission.
   \item  $\eta$ denotes the energy conversion efficiency, 
    \item  $P_{ps}$ denotes the transmission power of PS, 
      \item and $\zeta_1$ denotes the path loss from PS to U.
\end{itemize}

Following this, U transmits its information to BS via RIS during $(1-\alpha)\tau_c$ using the energy harvested during $\alpha\tau_c$. The transmission power of U can be calculated as,

 \begin{align}
 P_u= \frac{E}{(1-\alpha)\tau_c}=\frac{\eta  \alpha  P_{ps}  |h_1|^2 \zeta_1}{(1-\alpha)}=P_e|h_1 |^2 \zeta_1,
  \end{align}
  where $P_e=\frac{\eta  \alpha  P_{ps}}{(1-\alpha)}$

  At BS, the signal-to-noise ratio (SNR) can be written as,

    \begin{align}
   z_{u}=\frac{ P_e \zeta_1 \zeta_2 |h_1 |^2 |\textbf{h}_{3}^T \mathbf{\Theta} \textbf{h}_{2}|^2}{\sigma_{T}^2}=p_e \zeta_1 \zeta_2  |h_1 |^2 |\textbf{h}_{3}^T \mathbf{\Theta} \textbf{h}_{2} |^2
    \end{align}
  where 
  \begin{itemize}
    \item $\mathbf{\Theta}\!=\!\rho \diag  (e^{j\theta_1}, ..., e^{j\theta_{M}})$ is $M \times M$ diagonal response matrix of RIS
   \item $\rho$ is the amplification gain of each RIS
element\footnote{The amplification gain for each reflection-type amplifier is set to $\rho$, reducing configuration overhead and enabling variable beamforming design.}
\item $p_e=\frac{P_e}{\sigma_T^2}$, $\sigma_T^2=\sigma_b^2+\rho^2\sigma_R^2$ is sum of powers of AWGN noises at the BS and RIS, 
  \item and $\zeta_2$ is the path loss from $U$ to BS via RIS.
    \end{itemize}
 
All channels are assumed to be known \cite{b8}. This implies that the phase shifts at the RIS can be intelligently controlled, i.e.,  $\theta_m\!=-\!\arg \left(\textbf{h}_{2,m}  \textbf{h}_{3,m} \right)$, to maximize  the reception quality of BS \cite{b9}. In this case, the SNR in Eq. (3) can be rewritten as,

\begin{align}
z_{u}=p_e \zeta_1 \zeta_2 |h_1|^2\left(\rho \sum_{m=1}^{M}|\textbf{h}_{2,m}||\textbf{h}_{3,m} |\right)^2
\end{align}

\section{Outage Probability Analysis}

The outage probability is the probability of an event occurring when channel capacity drops below the codeword rate of the transmission signal ($r$), and is mathematically expressed as:

\begin{align}
&P_{Out}=\Pr\left[ \left(1-\alpha\right)\log_2\left(1+z_{u}\right)<r \right]
\end{align}

By substituting the SNR from Eq. (4), Eq. 5 can be accurately expressed as:

\begin{align}
P_{Out}&=\nonumber\\
&\Pr\left[ \left(1-\alpha\right)\log_2\left(p_e \zeta_1 \zeta_2 |h_1|^2\left(\rho \sum_{m=1}^{M}|\textbf{h}_{2,m}||\textbf{h}_{3,m} |\right)^2\right)<r\right] \nonumber\\
=&\Pr\left[\left(|h_1| \sum_{m=1}^{M}|\textbf{h}_{2,m}||\textbf{h}_{3,m} |\right)^2<\frac{\Delta}{p_e \zeta_1 \zeta_2 \rho^2}\right]
\end{align}

Using the statistical characteristics of Rayleigh channels and employing algebraic manipulations, Eq. (6) can be elegantly simplified as follows,

\begin{align}
P_{Out}=&1-\frac{\pi^2}{4K}\sum_{k=1}^{K}\sqrt{1-\omega_k^2}\sec^2\mu_k\frac{\left(\tan\mu_k\right)^{v-1}}{\Gamma(v)\delta^v} \nonumber\\
\times & \exp \left(-\frac{\Delta}{p_e \zeta_1 \zeta_2 \rho^2 tan^2\mu_k}-\frac{tan\mu_k}{\delta}\right)
\end{align}  
where 
\begin{itemize}
  \item $\mu_k=\frac{\pi\left(\omega_k+1\right)}{4}$,
    \item $\omega_k=\cos\left(\frac{2k-1}{K}\pi \right)$, 
      \item $K$ is the accuracy versus complexity parameter,
 \item  $\Delta=2^{\frac{r}{1-\alpha}}-1$, 
   \item $v=M\frac{\frac{\pi}{4}}{\frac{4}{\pi}-\frac{\pi}{4}}$,
     \item  $\delta=\frac{4}{\pi}-\frac{\pi}{4}$,   \item and $\Gamma (.)$ denotes the Gamma function.
     \end{itemize}

\section{Numerical Results}

Numerical results validate the theoretical expressions and provide valuable insights. Monte Carlo simulations with $10^4$ independent trials are used. Unless otherwise specified, the simulation parameters are as follows: $\alpha=0.5$, $\tau_c=1$, $\eta=0.90$, $r=1.4$ bps/Hz, $P_{ps}=10$ dBm, $M=\{0,100\}$, $\sigma_b^2=0$ dBm, and  $\sigma_R^2=-80$ dBm, and $\rho=\{0,10\}$ dB. 

Fig. 2 shows the outage probability  vs. number of active elements ($M$), and amlification factor for active elements ($\rho$). The results validate the analytical analysis and imply that an increase in $M$, and $\rho$ (which demonstrates the amplification impact due to the active RIS) positively impacts outage performance, subsequently increasing transmission reliability. As an example, deploying an active RIS rather than a passive RIS allows for a decrease in the transmit power as well as the number of elements required to achieve a certain outage probability.

\begin{figure}[t]
\centering
\includegraphics[width=3.4in,height=3in]{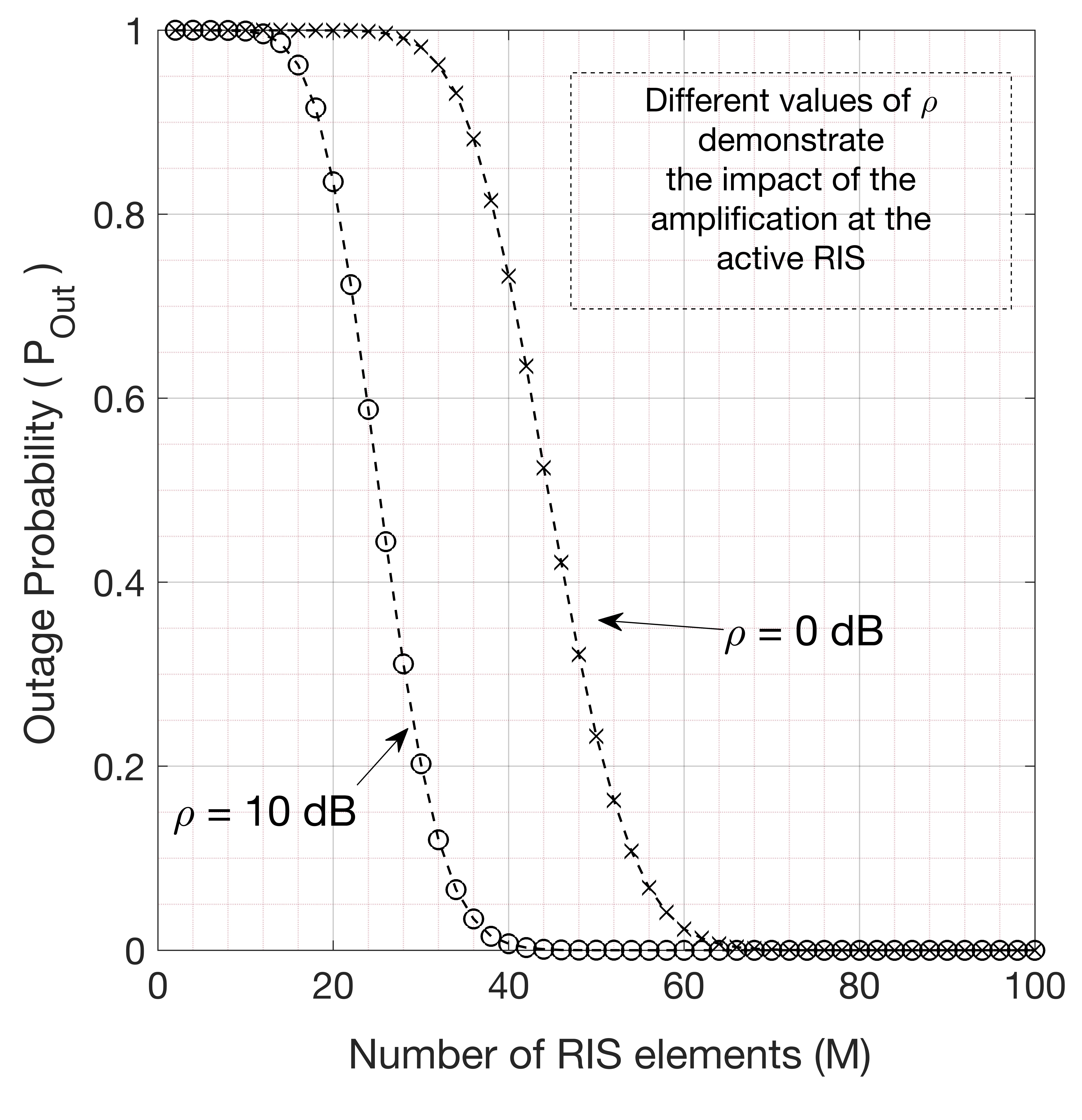}
\caption {Outage probability vs. Number of active elements ($M$) for different amplification gain ($\rho$)}
\label{diagram2}
\end{figure}

\section{Conclusion}

The integration of active RIS and WPC significantly enhances the reliability and energy efficiency of users with limited energy resources. This paper demonstrates how a user can harvest RF energy from a PS and utilize it for data transmission to a BS via RIS, optimizing the use of available energy and ensuring reliable communication. The theoretical analysis and numerical results provide crucial insights and practical guidelines for parameter selection, contributing to the advancement of energy-efficient and reliable wireless communication, particularly as we transition into the 6G era.

%




\begin{thebibliography}{00} 

\bibitem{b1} W. Khalid, {\em et al.}, ``Reconfigurable intelligent surface for physical layer security in 6G-IoT: Designs, issues, and advances," {\em IEEE Internet Things J.} (Early Access), doi: 10.1109/JIOT.2023.3297241.

\bibitem{b2} W. Khalid, {\em et al.}, ``Simultaneous transmitting and reflecting-reconfigurable intelligent surface in 6G: Design guidelines and future perspectives," {\em IEEE Netw.} (Early Access), doi: 10.1109/MNET.129.2200389.



\bibitem{b3} S. -M. Park, {\em et al.}, ``Joint antenna and device scheduling in full-duplex MIMO wireless-powered communication networks," {\em IEEE Internet Things J.}, vol. 9, no. 19, pp. 18908--18923, Oct. 2022.


\bibitem{b4} W. Khalid, {\em et al.}, ``Simultaneously transmitting and reflecting-reconfigurable intelligent surfaces with hardware impairment and phase error," {\em In Proc. 2023 ICAIIC}, Bali, Indonesia, 20-23 Feb. 2023.


\bibitem{b4a} W. Khalid, {\em et al.}, ``Impact of RIS on outage probability and ergodic Rate in wireless powered communication," {\em In Proc. 2023 ICTC}, Jeju, Korea, 11-13 Oct. 2023. DOI: 10.48550/arXiv.2401.01787.



\bibitem{b5} M. H. Khoshafa, {\em et al.}, ``Active reconfigurable intelligent surfaces-aided wireless communication system," {\em IEEE Commun. Lett.}, vol. 25, no. 11, pp. 3699--3703, Nov. 2021.



\bibitem{b6} T. Ji, {\em et al.}, ``Exploiting intelligent reflecting surface for enhancing full-duplex wireless-powered communication networks," {\em 	IEEE Trans. Commun.} (Early Access), doi: 10.1109/TCOMM.2023.3323532.



\bibitem{b7} W. Khalid and H. Yu, ``Security improvement with QoS provisioning using service priority and power allocation for NOMA-IoT networks," {\em IEEE Access}, vol. 9, pp. 9937--9948, Jan. 2021.


\bibitem{b8} W. Khalid, {\em et al.}, ``Rate-energy tradeoff analysis in RIS-SWIPT systems with hardware impairments and phase-based amplitude response," {\em IEEE Access}, vol. 10, pp. 31821--31835, Mar. 2022.

\bibitem{b9} W. Khalid, {\em et al.}, ``RIS-aided physical layer security with full-duplex jamming in underlay D2D networks," {\em IEEE Access}, vol. 9, pp. 99667--99679, Jul. 2021.





\normalsize


\end{thebibliography}
\end{document}